\documentclass{aastex631}

\shorttitle{X-ray Variability in AGNs} \shortauthors{Hu et al.}
\graphicspath{{./}{figures/}}
\usepackage{CJK}
\usepackage{amsmath}
\usepackage{longtable}
\usepackage{booktabs}
\usepackage{hyperref}
\usepackage[all]{hypcap}
\usepackage{CJK}
\makeatletter

\newcommand{\Rmnum}[1]{\expandafter\@slowromancap\romannumeral #1@}
\makeatother

\begin{document}
\begin{CJK*}{UTF8}{gbsn}

\title{Magnetic Reconnection as a Potential Driver of X-ray Variability in Active Galactic Nuclei}

\author[0000-0002-5238-8997]{Chen-Ran Hu (胡宸然)}
\affiliation{School of Astronomy and Space Science, Nanjing
University, Nanjing 210023, China}

\author[0000-0001-7199-2906]{Yong-Feng Huang (黄永锋)}
\thanks{Email: hyf@nju.edu.cn}
\affiliation{School of Astronomy and Space Science, Nanjing
University, Nanjing 210023, China}
 \affiliation{Key Laboratory of
Modern Astronomy and Astrophysics (Nanjing University), Ministry
of Education, China}

\author[0000-0003-0721-5509]{Lang Cui (崔朗)}
\affiliation{Xinjiang Astronomical Observatory, Chinese Academy of
Sciences, 150 Science 1-Street, Urumqi 830011, China}
\affiliation{Key Laboratory of Radio Astronomy and Technology,
Chinese Academy of Sciences,  A20 Datun Road, Chaoyang District,
Beijing 100101, China}
 \affiliation{Xinjiang Key Laboratory of
Radio Astrophysics, 150 Science 1-Street, Urumqi 830011, China}

\author[0009-0006-5995-5225]{Hanle Zhang (张晗乐)}
\affiliation{School of Astronomy and Space Science, Nanjing
University, Nanjing 210023, China}

\author[0000-0001-6239-3821]{Jiang-Tao Li (李江涛)}
\affiliation{Purple Mountain Observatory, Chinese Academy of
Sciences, Nanjing 210023, China}

\author[0000-0001-7500-0660]{Li Ji (纪丽)}
\thanks{Email: ji@pmo.ac.cn}
\affiliation{Purple Mountain Observatory, Chinese Academy of
Sciences,  Nanjing 210023, China}
 \affiliation{Key Laboratory of
Dark Matter and Space Astronomy, Chinese Academy of Sciences,
Nanjing 210023, China}

\author[0000-0001-9648-7295]{Jin-Jun Geng (耿金军)}
\affiliation{Purple Mountain Observatory, Chinese Academy of
Sciences,  Nanjing 210023, China}

\author[0000-0003-3230-7587]{Orkash Amat (吾热卡西·艾麦提)}
\affiliation{School of Astronomy and Space Science, Nanjing
University,  Nanjing 210023,  China}

\author[0000-0001-7943-4685]{Fan Xu (许帆)}
\affiliation{Department of Physics, Anhui Normal University,
Wuhu 241002,  China}

\author[0009-0002-8460-1649]{Chen Du (杜琛)}
\affiliation{School of Astronomy and Space Science, Nanjing
University, Nanjing 210023,  China}

\author[0009-0008-6247-0645]{Wen-Long Zhang (张文龙)}
\affiliation{Purple Mountain Observatory, Chinese Academy of
Sciences,  Nanjing 210023, China}
 \affiliation{School of Astronomy and Space Sciences, University of
Science and Technology of China, Hefei 230026, China}

\author[0000-0002-6189-8307]{Ze-Cheng Zou (邹泽城)}
\affiliation{School of Astronomy and Space Science, Nanjing
University, Nanjing 210023, China}

\author[0009-0000-0467-0050]{Xiao-Fei Dong (董小飞)}
\affiliation{School of Astronomy and Space Science, Nanjing
University, Nanjing 210023,  China}

\author[0000-0002-2191-7286]{Chen Deng (邓晨)}
\affiliation{School of Astronomy and Space Science, Nanjing
University, Nanjing 210023, China}

\author[0000-0003-3166-5657]{Pengfei Jiang (蒋鹏飞)}
\affiliation{Xinjiang Astronomical Observatory, Chinese Academy of
Sciences, 150 Science 1-Street, Urumqi 830011, China}
\affiliation{Key Laboratory of Modern Astronomy and Astrophysics
(Nanjing University), Ministry of Education, China}

\author[0009-0007-7562-7720]{Jie Liao (廖杰)}
\affiliation{Xinjiang Astronomical Observatory, Chinese Academy of
Sciences, 150 Science 1-Street, Urumqi 830011, China}
\affiliation{College of Astronomy and Space Science, University of
Chinese Academy of Sciences, No.1 Yanqihu East Road, Beijing
101408, China}

\begin{abstract}

We present a systematic analysis on the X-ray variability in 13
bright quasars at $z > 4.5$, combining recent Swift observations
from 2021 to 2023 and archival multi-epoch observations. Upper
limits of the luminosity measurements were included in the
analysis by using the Kaplan-Meier estimator method. It is
found that the high-redshift quasars exhibit X-ray variability on
both short-term (hours-to-days) and intermediate-term
(weeks-to-months) timescales, with short-term variability
dominating the overall variation. A linear correlation exists
between the global mean ($\mu_{L_{\rm{2\!\text{--}\!10keV}}}$) and
standard deviation ($\sigma_{L_{\rm{2\!\text{--}\!10keV}}}$) of
X-ray luminosities, which is independent of the X-ray photon index
and optical-to-X-ray spectral slope. The localized stochastic
magnetic reconnection mechanism is strongly favored, which can
naturally lead to a scale-invariant power-law energy distribution
and satisfactorily explain the correlation. The $\sigma$-$\mu$
correlation parallels with the well-documented rms-flux relation
of low-redshift active galactic nuclei (AGNs), implying that the
magnetic reconnection mechanism could drive short-timescale X-ray
variability in both high- and low-redshift AGNs. The
highest-redshift quasar in our sample, J142952+544717 ($z =
6.18$), shows a luminosity distribution extending to ${10}^{47}\
\rm{erg\ {s}^{-1}}$ with a not conspicuous median luminosity. On
the other hand, J143023+420436 ($z = 4.7$), which hosts the most
relativistic jet among known high-redshift blazars, is dominated
in the high-luminosity regime (${10}^{47}\ \rm{erg\ {s}^{-1}}$ ),
making it an ideal target for multi-wavelength follow-up
observations. J090630+693030 is found to have a rest-frame period
of 182.46 days and J143023+420436 has a period of 16.89 days,
both could be explained by the global evolution of plasmoid
chains, in which magnetic islands formed during reconnection may
merge successively.

\end{abstract}


%----------------------------------------------------------------------
%                               Introduction
%----------------------------------------------------------------------

\section{Introduction}
\label{sec1:introduction}

Active galactic nuclei (AGNs) are highly luminous galaxy cores
powered by accretion onto a central supermassive black hole
(SMBH), with multi-wavelength emission spanning from radio to
gamma-rays. Their energy release mainly comes from gravitational
energy, which constitutes a distinct class of baryonic energy
reservoirs in the cosmos that differs from stellar nucleosynthesis
and matter annihilation processes. As a luminous subclass of AGNs,
quasars outshine their host galaxy by orders of magnitude, with
their emission coming from an extremely compact region ($<$1 pc).
Their extraordinary brightness makes them visible even at extreme
cosmological distances, with current record-holding quasars
observed at redshifts  $z \approx 7.5\text{--}7.6$
\citep{2020ApJ...897L..14Y, 2021ApJ...907L...1W}. The unique
brightness establishes quasars as invaluable cosmological probes
\citep{2022MNRAS.513.5686C, 2022MNRAS.515.1795B,
2022ApJ...931..106D}, which have been utilized as standard candles
for cosmological distance measurements to explore the expansion
history of the universe \citep{2017A&A...602A..79L,
2019A&A...631A.120S, 2020A&A...642A.150L}.

As bright persistent objects in the early universe, quasars also
serve critical roles in investigating primordial structure
formation \citep{2006NewAR..50..140D, 2021PhRvL.126a1101K} and
cosmic reionization history \citep{2001AJ....122.2850B,
2006AJ....132..117F, 2015ApJ...813L...8M, 2015A&A...578A..83G}.
Moreover, the complex interplay between SMBH accretion and AGN
feedback plays a key role in regulating the co-evolution of AGN
and their host galaxies \citep{1998A&A...331L...1S,
2009Natur.460..213C, 2013ARA&A..51..511K, 2020MNRAS.493.1888T,
2022MNRAS.512.1052P}. A systematic investigation of quasar
evolution from cosmic dawn to present epochs therefore provides
critical insights into SMBH growth \citep{2005Natur.435..629S,
2024ApJ...964...39G} and galaxy evolution
\citep{2017A&A...601A.143F, 2024Natur.632.1009W}.

X-ray emission functions as a significant feature of AGN and is
pivotal for multiwavelength studies \citep{2020ApJ...892..103Z,
2020A&A...635L...7B, 2021ApJ...908..157P}. The advantage of X-ray
observations lies in their relative insensitivity to circumnuclear
obscuration and capacity to provide comprehensive diagnostics,
making them particularly useful for AGN identification in
large-scale surveys \citep{2011ApJS..195...10X,
2017ApJS..228....2L}.

The X-ray emission of AGN may be produced via multiple mechanisms.
The predominant mechanism involves inverse Compton scattering
(ICS), where optical/ultraviolet (UV) photons from the accretion
disk are up-scattered by thermal electrons in the corona
\citep{1991ApJ...380L..51H, 1993ApJ...413..507H,
1994MNRAS.269L..55Z, 2000ApJ...542..703Z, 2009ApJ...693..174C,
2012MNRAS.424.1284W}. Relativistic jets collimated perpendicular
to the accretion disk \citep{1993ARA&A..31..473A,
1995PASP..107..803U} can also lead to X-ray emission through
synchrotron self-Compton (SSC) processes
\citep{2005ApJ...622..797K}. While such a jet emission is
typically prominent in radio bands \citep{2012ARA&A..50..455F}, it
can also be detected in optical \citep{2021ApJ...913..146M} and
X-ray regimes \citep{2016MNRAS.455.3526H}. Thermal radiation from
the accretion disk is also mainly in X-rays, which may explain the
UV/soft X-ray excess observed in some Seyfert-1 galaxies
\citep{2004MNRAS.349L...7G}, although competing interpretations
have also been proposed \citep{2006MNRAS.365.1067C,
2012MNRAS.420.1848D}. Fluorescent line emission generated through
photoionization of broad-line region clouds
\citep{2007MNRAS.382..194N} and reprocessed radiation from the
obscuring torus \citep{2005A&A...444..119G, 2009MNRAS.397.1549M}
can also contribute X-ray emission. Additionally, other X-ray
emission mechanisms include shock-heated thermal radiation
produced by AGN-driven winds \citep{2001ApJ...550..230L,
2012MNRAS.425..605F, 2015Natur.519..436T, 2021NatAs...5..928S},
transient emission arising from tidal disruption events of
SMBH-star systems \citep{2011Natur.476..421B,
2017ApJ...835..144G}, and hydrodynamic interactions during the
spiral-in of binary AGN systems.

Note that the jet of most AGNs is misaligned with our line of
sight. As a result, the coronal ICS process becomes the dominant
X-ray emission mechanism. However, in blazars, the AGN subclass
with a relativistic jet oriented toward us, a substantial fraction
of X-ray emission arises from SSC radiation within the jet
\citep{1985ApJ...298..114M, 2005ApJ...622..797K,
2014MNRAS.439..690H, 2019NewAR..8701541H}. Observational selection
effects likely enhance the apparent blazar fraction among
high-redshift quasars \citep{2014ApJ...780...73A}, introducing
further complexity in interpreting the X-ray variation patterns of
high-redshift AGNs.

AGNs exhibit variability at all wavelengths, with temporal scales
ranging from minutes to years \citep{2004ApJ...605..662C,
2005ApJ...622..160X, 2008A&A...485...51H, 2009ApJ...705..199B,
2010MNRAS.401.1570T, 2018ApJ...853...34Z}. X-ray variability in
AGNs not only serves as an identification tool
\citep{2012ApJ...748..124Y} but also provides effective probes for
understanding the underlying physics. It can help constrain the
SMBH spin \citep{2016MNRAS.461.1967I}, accretion physics
\citep{2006Natur.444..730M}, corona geometry and properties
\citep{2015MNRAS.448..703W, 2015MNRAS.449..129W,
2015MNRAS.454.4440W}, jet dynamics and emission
\citep{2010ApJ...710L.126M}, SMBH binary candidates
\citep{2009ApJ...696.2170R, 2014ApJ...786..103L}, and even provide
useful clues for understanding the Galactic black hole X-ray
binaries (BHXRBs) \citep{2012A&A...544A..80G}. X-ray variability
can also help constrain SMBH masses \citep{2001ApJ...550L..15P,
2006Natur.444..730M}, which is a useful supplement to the popular
reverberation mapping method for mass measurement
\citep{1993PASP..105..247P, 2004ApJ...613..682P}.

X-ray variability of low-to-moderate redshift AGNs have been
extensively studied \citep{2012A&A...544A..80G,
2017A&A...599A..82M, 2020MNRAS.498.4033T, 2021ApJ...910..103L}.
However, systematic investigation of X-ray variability of AGNs at
high redshift still remains scarce. Recently, the number of
high-redshift AGNs ($z > 4.5$) detected in X-rays is increasing
due to advances in observational capability
\citep{2021MNRAS.504.2767L, 2021ApJ...906..135L}. In this study,
we present a comprehensive analysis on the X-ray variability of 13
bright X-ray quasars lying at $z > 4.5$, taking into account
recent new Swift follow-up observations on them. Potential
mechanisms driving the X-ray variability in high-redshift AGNs are
explored.

%----------------------------------------------------------------------
%                 Data Reduction and Analytical Methods
%----------------------------------------------------------------------

\section{Data Reduction and Analysis Method}
\label{sec2:data reduction and analytical methods}

\subsection{Swift Observations and Data Reduction}
\label{sec2.1:Swift observations and data reduction}

X-ray surveys conducted with the Chandra X-ray Observatory and the
X-Ray Multi-Mirror Mission Observatory (XMM-Newton) have led to
the detection of emission from over 100 quasars at $z \geq 4.5$
\citep{2021MNRAS.504.2767L, 2021ApJ...906..135L}. From a broader
sample of high-redshift quasars, we selected 13 bright sources at
$z \geq 4.5$ with new multi-cycle follow-up
observations by the X-Ray Telescope (XRT) on
board the Neil Gehrels Swift Observatory (Swift) from 2021 to
2023. Their key parameters are summarized in Table \ref{table1}.
For these 13 quasars, we performed a detailed analysis by
combining new Swift observations with
archival Swift data
(\href{https://heasarc.gsfc.nasa.gov/cgi-bin/W3Browse/swift.pl}
{https://heasarc.gsfc.nasa.gov/cgi-bin/W3Browse/swift.pl}).
Previous X-ray observational data from other missions such as
Chandra, XMM-Newton, the extended ROentgen Survey with an Imaging
Telescope Array (eROSITA), and the Nuclear Spectroscopic Telescope
Array (NuSTAR) were also included in our study (see Table
\ref{table1}).

The Swift data (Level 2 FITS files, i.e., calibrated event lists) were
processed using HEASoft (v6.30). The background-subtracted net
counts of the sources were extracted utilizing the standard
on-source/off-source technique. For each source, a circular source region
(radius = $30\arcsec$) centered on the source position was defined, while an annular
background region (inner/outer radii = $30\arcsec$/$480\arcsec$) was
selected to exclude contamination. The background contribution was scaled by
the area ratio between the source and background regions before subtraction.

In Swift XRT's photon counting mode, the telescope exposure time ranges
from several hundred seconds to several thousand seconds
during grazing scans of sources. Due to the limited photon counts from these
high-redshift sources caused by short exposure times, we adopted the Li-Ma
formula \citep{1983ApJ...272..317L}, a standard significance evaluation
method in gamma-ray astronomy, as a replacement for conventional
signal-to-noise ratio calculations. This approach mitigates systematic underestimation
of faint signal significance. To enhance signal significance to above
$3\sigma$, we merged temporally adjacent observations with rest-frame
separations $\leq3$ days. To maximize the number of photometric
data points, we optimized merging observations to achieve merged
significances marginally exceeding $3\sigma$. Adjacent observations
within 3 days (rest-frame) whose combined significance remained below
$3\sigma$ were conservatively treated as upper limits
 (see Section \ref{sec2.2:Kaplan-Meier estimator}).

\clearpage

%% Table1: Basic Properties of 13 X-ray Brightest Quasars at $z > 4.5$
\begin{longrotatetable}
\begin{deluxetable*}{ccccccccccccc}
\renewcommand{\arraystretch}{1.8}
\tabletypesize{\scriptsize} \tabcolsep=3.5pt
\tablecaption{
Basic parameters of the 13 bright quasars at $z > 4.5$
}
\label{table1} \tablehead{
\colhead{Quasar Name} & \colhead{RA} & \colhead{DEC} &
\colhead{$z$} & \colhead{$\Gamma$} & \colhead{$N_{\rm{H}}$} &
\colhead{$\alpha_{\rm{OX}}$} & \colhead{$M_{1450}$}& \colhead{$L_{2500}$} &
\colhead{$m_{\rm{SMBH}}$} & \colhead{$\lambda_{\rm{Edd}}$} &
\colhead{Note} & \colhead{Ref.}\\
\colhead{} & \colhead{} & \colhead{} & \colhead{} & \colhead{} &
\colhead{($10^{20}\,{\rm{cm}}^{-2}$)} & \colhead{} & \colhead{} &
\colhead{($10^{32}\,{\rm{erg}}\,{\rm{s}}^{-1}\,{\rm{Hz}}^{-1}$)} &
\colhead{($10^{10}\,M_\odot$)} & \colhead{} & \colhead{} & \colhead{}
}
\startdata
    J001115+144601 &  00:11:15.235  &  +14:46:01.80  & 4.96  & $1.78^{+0.15} _{-0.15}$ & 5.00  & $1.24^{+0.01}_{-0.01}$ & -28.03  & 0.93  &       &       &       & (1) \\
    J013127-032100 &  01:31:27.34  &  -03:21:00.1  & 5.18  & $1.89^{+0.07}_{-0.07}$ & 11.00  &       &       &       &       &       & Blazar & (2) \\
    J032444-291821 &  03:24:44.293  &  -29:18:21.24  & 4.62  & $1.62^{+0.20}_{-0.20}$ & 5.00  & $1.43^{+0.02}_{-0.02}$ & -28.26  & 1.15  &       &       & Radio-loud blazar & (1), (3), (4) \\
    J035504-381142 &  03:55:04.893  &  -38:11:42.51  & 4.58  & $1.71^{+0.26}_{-0.26}$ & 5.00  & $1.50^{+0.02}_{-0.02}$ & -28.56  & 1.51  &       &       &       & (1) \\
    J074749+115342 &  07:47:49.202  &  +11:53:52.90  & 5.26  & $2.07^{+0.17}_{-0.17}$ & 5.00  & $1.46^{+0.02}_{-0.02}$ & -28.04  & 0.94  & 0.18  & $2.25^{+0.09}_{-0.09}$ & Radio-quiet & (1), (5) \\
    J090630+693030 &  09:06:30.773  &  +69:30:30.65  & 5.47  & $1.50^{+0.07}_{-0.07}$ & 5.00  & $1.13^{+0.01}_{-0.01}$ & -26.72  & 0.28  & 0.42  &       & Radio-loud blazar & (1), (6) \\
    J094004+052630 &  09:40:04.827  &  +05:26:30.43  & 4.5   & $2.01^{+0.32}_{-0.31}$ & 5.00  & $1.08^{+0.04}_{-0.04}$ & -25.29  & 0.07  &       &       & Radio-loud & (1), (7), (8) \\
    J102623+254259 &  10:26:23.628  &  +25:42:59.82  & 5.25  & $1.31^{+0.23}_{-0.23}$ & 5.00  & $1.18^{+0.03}_{-0.03}$ & -26.69  & 0.27  &       &       &       & (1) \\
    J142952+544717 &  14:29:52.146  &  +54:47:17.55  & 6.18  & $1.40^{+0.90}_{-0.90}$ & 1.15  & $1.12^{+0.13}_{-0.57}$ &       & 0.16  & $\geq 0.15$ &       & Radio-loud & (9), (10), (11) \\
    J143023+420436 &  14:30:23.745  &  +42:04:36.62  & 4.7   & $1.35^{+0.03}_{-0.03}$ & 5.00  & $0.70^{+0.004}_{-0.004}$ & -26.39  & 0.20  &       &       & Radio and gamma-ray blazar & (1), (12), (13) \\
    J145147-151220 &  14:51:47.054  &  -15:12:20.15  & 4.76  & $1.62^{+0.16}_{-0.16}$ & 5.00  & $1.44^{+0.02}_{-0.02}$ & -28.98  & 2.24  &       &       &       & (1) \\
    J154824+333500 &  15:48:24.027  &  +33:35:00.53  & 4.68  & $2.15^{+0.22}_{-0.22}$ & 5.00  & $0.97^{+0.03}_{-0.03}$ & -25.42  & 0.08  &       &       & Radio-loud & (1), (7), (8) \\
    J210240+601509 &  21:02:40.225  &  +60:15:09.64  & 4.57  & $1.15^{+0.30}_{-0.30}$ & 5.00  & $0.95^{+0.05}_{-0.04}$ & -25.06  & 0.06  &       &       &       & (1) \\
\enddata
\tablecomments{
The data were collected as of 2024 October.\\
Columns 2--13: right ascension (J2000) in units of degree; declination
(J2000) in units of degree; redshift; X-ray photon index; hydrogen column
density; optical-to-X-ray spectral slope; Absolute magnitude at 1450 \text{\AA};
rest-frame 2500 \text{\AA} monochromatic luminosity; SMBH mass; Eddington ratio of
SMBH; additional description; references.\\
References: (1) \citet{2021MNRAS.504.2767L};
(2) \citet{2020ApJ...904...27A}; (3) \citet{2014ApJ...795L..29Y};
(4) \citet{2015MNRAS.450L..57G}; (5) \citet{2021ApJ...906..135L};
(6) \citet{2006AJ....132.1959R}; (7) \citet{2016MNRAS.463.3260C};
(8) \citet{2020ApJ...899..127S}; (9) \citet{2020MNRAS.497.1842M};
(10) \citet{2021MNRAS.504..576M}; (11) \citet{2023MNRAS.524.1087M}
(12) \citet{2020SciBu..65..525Z}; (13) \citet{2018ApJ...865L..17L}.\\
}
\end{deluxetable*}
\end{longrotatetable}

For observations with insufficient photon counts to perform reliable
spectral fits, we derived the absorption-corrected energy flux from
the observed net photon flux in the 0.5 -- 2 keV band with a constant
conversion factor by assuming an absorbed power-law spectral model. The
photon-to-energy flux conversion factor was computed using the HEASARC
WebPIMMS tool, with model parameters (hydrogen column density $N_{\rm{H}}$,
photon index $\Gamma$, and redshift $z$) adopted from the literature \citep{2021MNRAS.504.2767L,
2020ApJ...904...27A, 2020MNRAS.497.1842M}. The rest-frame
absorption-corrected 2 -- 10 keV energy flux can then be obtained after
correcting for the redshift effect.

\subsection{Kaplan-Meier Estimator Method}
\label{sec2.2:Kaplan-Meier estimator}

Due to the limited exposure time, only an upper flux limit could be
derived in many Swift-XRT observations. In most cases, such upper
limits are simply omitted in quantitative analysis. Occasionally,
people try to incorporate the upper limits through modeling and Monte Carlo
simulations, but such approaches generally depend on implicit assumptions
about the nature of the upper limits.
Given the scarcity of high-redshift AGN observations, effectively utilizing
these upper limits becomes critical.

The Kaplan-Meier estimator method, a classical survival analysis tool widely
applied in biomedical research, offers a robust solution for data with
upper/lower limits \citep{kaplan1958, 1985ApJ...293..192F}. In clinical trials tracking
patient relapse times, subjects lost to follow-up without recurrence provide
only lower-bound measurements for time to
recurrence. Kaplan \& Meier (1958) addressed this by treating
both relapse events (firm observations) and censoring instances (lower
limits, i.e., subjects lost to follow-up without recurrence) as ``event time
nodes'', assuming: (1) independence between all event times, and (2)
identical distributions for censoring events (lower limits). This weak
assumption framework enables non-parametric estimation of relapse time
distributions without prior distribution assumptions.

In our study, we have adopted the Kaplan-Meier estimator method to deal with
the observational upper limits. For this purpose, we need to extend the
formulas correspondingly. Let $T$ denote the random variable for relapse time $t$.
The cumulative distribution function (CDF) $F(t)$ and the survival function
$S(t)$ are
\begin{eqnarray}
\label{formula1}
F(t)=P(T\le t) ,
\end{eqnarray}
\begin{eqnarray}
\label{formula2}
S(t)=P(T\geq t)=1-F(t) .
\end{eqnarray}
For $n$ observed event times $\left\{\tau_i\right\}_{i=1}^n$ (including both
relapses and censoring), let the distinct ordered event times be
$\tau^\prime_{(1)}<\tau^\prime_{(2)}<\ldots<\tau^\prime_{(r)}$,
with $\tau^\prime_{(0)}=0$. For $j=0,1\ldots,r-1$, the conditional probability
is denoted as
\begin{eqnarray}
\label{formula3}
P_j=P\left[T\geq\left.\tau^\prime_{(j+1)}\right|T\geq\tau^\prime_{(j)}\right] .
\end{eqnarray}

According to the Kaplan-Meier estimator method, the non-parametric maximum
likelihood function based on the probability of discrete events is
\begin{eqnarray}
\label{formula4}
L=\prod_{j}(1-P_j)^{d_j}{P_j}^{n_j-d_j} ,
\end{eqnarray}
where
\begin{eqnarray}
\label{formula5}
d_j=\#\left\{k,\tau_k=\tau^\prime_{(j)}\right\} ,
\end{eqnarray}
\begin{eqnarray}
\label{formula6}
n_j=\#\left\{k,\tau_k\geq\tau^\prime_{(j)}\right\} .
\end{eqnarray}
At each distinct event time $\tau^\prime_{(j)}$, $d_j$
corresponds to the number of subjects experiencing relapse (firm
observations), while $n_j$ denotes the number of subjects remaining
relapse-free and under follow-up. Based on Equation (\ref{formula4}), the
maximizing solution for $P_j$ is derived as
\begin{eqnarray}
\label{formula7}
\hat{P}_j=1-\frac{d_j}{n_j} .
\end{eqnarray}

Substituting Equations (\ref{formula3}) and (\ref{formula7}) into Equation
(\ref{formula2}), we obtain
\begin{eqnarray}
\label{formula8}
\hat{S}(t)=
\left\{
    \begin{array}{lc}
         1, \ \ 0\le t\le\tau^\prime_{(1)} ,  \\
         \prod_{j}(1-\frac{d_j}{n_j}), \ \ \tau^\prime_{(j)}\le t\le\tau^\prime_{(j+1)},\,j>0 .
    \end{array}
\right.
\end{eqnarray}
Note that to ensure $\lim\limits_{t\to\infty} S(t)=0$, the
maximum observed event time $\tau^\prime_{(r)}$ must be treated
as a lower limit, even if it is not. Subsequently, the mean $\mu_t$ and
variance $\sigma_t^2$ of the event times $t$ can be derived as
\begin{eqnarray}
\label{formula9}
\mu_t=\int_{0}^{\infty}tf(t){\rm{d}}t=-\int_{1}^{0}t{\rm{d}}S(t)=\int_{0}^{\infty}S(t){\rm{d}}t-\left[tS(t)\right]_0^\infty=\int_{0}^{\infty}S(t){\rm{d}}t ,
\end{eqnarray}
\begin{eqnarray}
\label{formula10}
\sigma_t^2=\int_{0}^{\infty}{(t-\mu_t)^2f(t){\rm{d}}t}=\int_{0}^{\infty}{t^2f(t){\rm{d}}t}-\mu_t^2=\left[t^2S(t)\right]_0^\infty-\int_{0}^{\infty}2tS(t){\rm{d}}t-\mu_t^2=2\int_{0}^{\infty}tS(t){\rm{d}}t-\mu_t^2 .
\end{eqnarray}
Here, $f(t)$ denotes the probability density function of event
times $t$. The median event time $M_t$, defined by $S(M_t)=0.5$,
can be directly obtained from the survival function. For discrete
observational data, the estimators in Equations (\ref{formula9}) and
(\ref{formula10}) are approximated as
\begin{eqnarray}
\label{formula11}
\begin{aligned}
\hat{\mu}_t & =\frac{\tau^\prime_{(1)}-\tau^\prime_{(0)}}{2}\hat{S}[\tau^\prime_{(0)}]+\frac{\tau^\prime_{(r)}-\tau^\prime_{(r-1)}}{2}\hat{S}[\tau^\prime_{(r)}]+\sum_{j=1}^{r-1}{\frac{\tau^\prime_{(j+1)}-\tau^\prime_{(j-1)}}{2}\hat{S}[\tau^\prime_{(j)}]} \\
& =\frac{\tau^\prime_{(1)}}{2}+\sum_{j=1}^{r-1}{\frac{\tau^\prime_{(j+1)}-\tau^\prime_{(j-1)}}{2}\hat{S}[\tau^\prime_{(j)}]} ,
\end{aligned}
\end{eqnarray}
\begin{eqnarray}
\label{formula12}
\begin{aligned}
\hat{\sigma}_t^2 & =\left[\tau^\prime_{(1)}-\tau^\prime_{(0)}\right]\tau^\prime_{(0)}\hat{S}[\tau^\prime_{(0)}]+\left[\tau^\prime_{(r)}-\tau^\prime_{(r-1)}\right]\tau^\prime_{(r)}\hat{S}[\tau^\prime_{(r)}]+\sum_{j=1}^{r-1}{\left[\tau^\prime_{(j+1)}-\tau^\prime_{(j-1)}\right]\tau^\prime_{(j)}\hat{S}[\tau^\prime_{(j)}]}-\hat{\mu}_t^2 \\
& =\sum_{j=1}^{r-1}{\left[\tau^\prime_{(j+1)}-\tau^\prime_{(j-1)}\right]\tau^\prime_{(j)}\hat{S}[\tau^\prime_{(j)}]}-\hat{\mu}_t^2 .
\end{aligned}
\end{eqnarray}
Here, the integration can be handled by employing the trapezoidal rule.

The Kaplan-Meier estimator method was originally developed for data analysis with lower
limits in biomedical studies, therefore necessary modifications are required to
accommodate the astronomical data with upper limits. For an astronomical
dataset $\left\{\alpha_i\right\}_{i=1}^n$  (corresponding to random variable
$A$ with values $a$), we apply the following transformation to convert it
into data with lower limits:
\begin{eqnarray}
\label{formula13}
\tau_i=\max\left\{\alpha_i\right\}_{i=1}^n-\alpha_i .
\end{eqnarray}
The transformed values $\left\{\tau_i\right\}_{i=1}^n$ now are compatible
with Equations (\ref{formula1})--(\ref{formula12}) while preserving the
information encoded in the original dataset $\left\{a_i\right\}_{i=1}^n$.
Accordingly, we can obtain:
\begin{eqnarray}
\label{formula14}
F(a)=P(A\le a)=P(\max\left\{\alpha_i\right\}_{i=1}^n-T\le a)=P(T\geq\max\left\{\alpha_i\right\}_{i=1}^n-a)=P(T\geq t)=S(t) ,
\end{eqnarray}
\begin{eqnarray}
\label{formula15}
M_a=S_a^{-1}(0.5)=\max\left\{\alpha_i\right\}_{i=1}^n-S_t^{-1}(0.5)=\max\left\{\alpha_i\right\}_{i=1}^n-M_t ,
\end{eqnarray}
\begin{eqnarray}
\label{formula16}
\hat{\mu}_a=E(a)=E(\max\left\{\alpha_i\right\}_{i=1}^n-t)=\max\left\{\alpha_i\right\}_{i=1}^n-E(t)=\max\left\{\alpha_i\right\}_{i=1}^n-\hat{\mu}_t ,
\end{eqnarray}
\begin{eqnarray}
\label{formula17}
\hat{\sigma}_a^2= D(a)=D(\max\left\{\alpha_i\right\}_{i=1}^n-t)=D(-t)=D(t)=\hat{\sigma}_t^2 ,
\end{eqnarray}
where $E(a)$ and $D(a)$ denote the expectation
and variance of $a$, respectively.

Using Equations (\ref{formula14})--(\ref{formula17}), we can
perform a robust statistical analysis on the astronomical data, taking into
account those upper limits with minimal assumptions. For large samples, $\hat{\mu}_a$
and $\hat{\sigma}_a^2$ asymptotically follow normal distributions so that their
errors can be easily derived \citep{1985ApJ...293..192F}. In small size samples,
Monte Carlo simulations will be used to estimate the uncertainties by varying the firm
detections within their error ranges while preserving the upper limits.

%----------------------------------------------------------------------
%      X-ray Variability of 13 X-ray Brightest Quasars at $z > 4.5$
%----------------------------------------------------------------------

\section{X-ray Variability of the High Redshift Quasars}
\label{sec3:X-ray variability of 13 X-ray brightest quasars at $z > 4.5$}

\begin{figure}[htbp]
\gridline{\fig{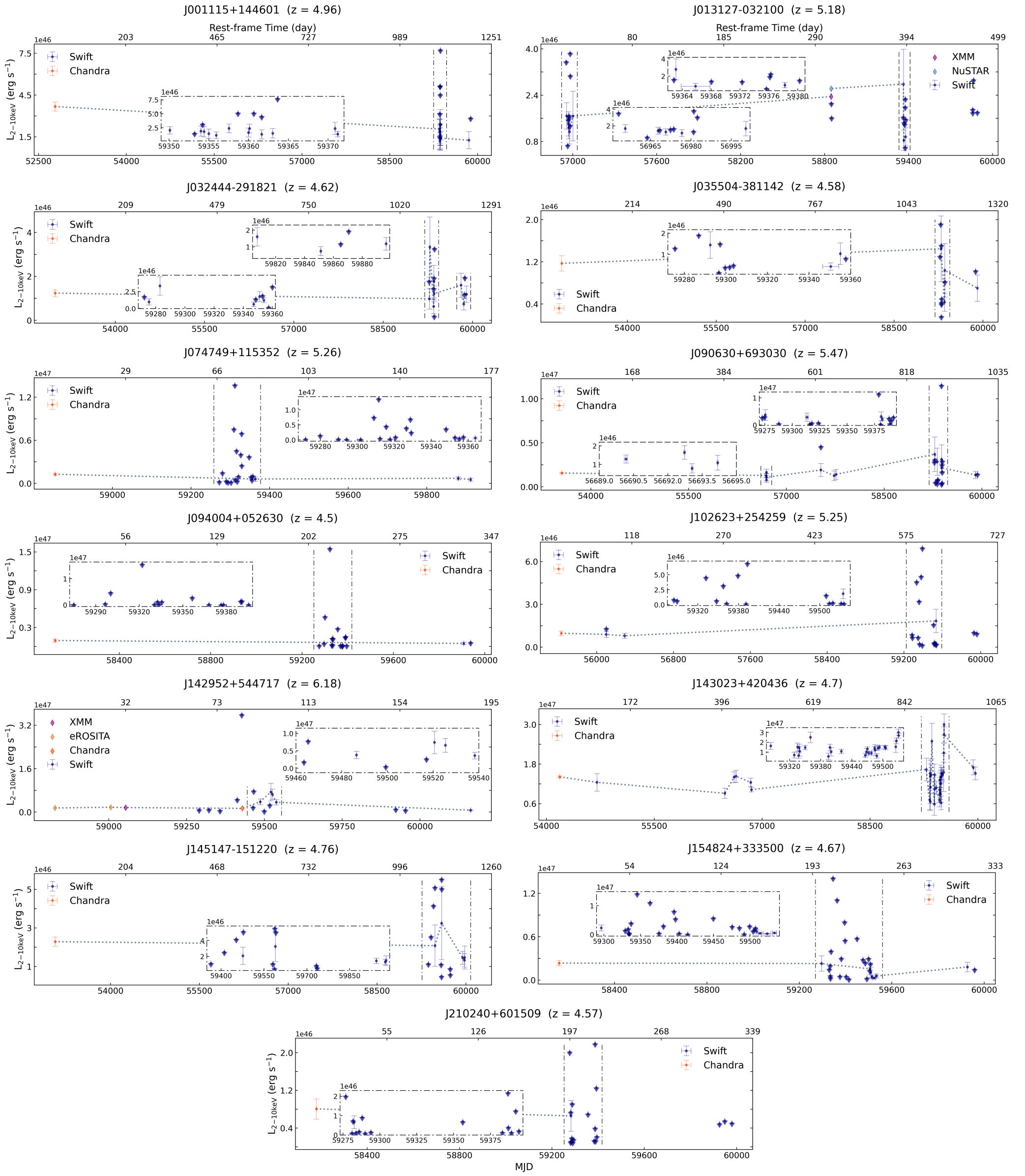}{1\textwidth}{}}
\caption{
X-ray light curves of the 13 bright quasars at $z > 4.5$.
The error bars of the observational data points correspond to
$1\sigma$ uncertainties, while the upper limits are shown at the $3\sigma$ confidence level.
In each panel, the rectangular inset provides a zoomed-in view of the region bounded by the
corresponding line style.
}
\label{figure1}
\end{figure}

X-ray light curves of the 13 bright quasars at $z > 4.5$ are shown
in Figure \ref{figure1}, which includes
Swift observations as well as archival data from various X-ray
telescopes (see Table \ref{table1} for data references). Most of
the sources were regularly monitored by Chandra and Swift, but
note that two sources are slightly different. J013127-032100,
lying at $z = 5.18$, were frequently observed by XMM, NuSTAR, and
Swift. The most distant source in our sample, J142952+544717 ($z =
6.18$) were monitored by eROSITA, XMM, Chandra, and Swift.

Most sources have been monitored for 1 -- 3.5 years (rest-frame time), but
J074749+115342 and J142952+544717 were monitored for less than $\sim0.5$ years.
These quasars show apparent variability in their brightness, with the varying
timescale ranging from short-term (hours-to-days) to intermediate-term (weeks-to-months)
in the rest frame. The amplitude of luminosity variation generally remains
in one order of magnitude across all timescales if the upper limits are not considered.
Here we perform a more in-depth analysis on the variability.
To mitigate the selection biases involving upper limits, the Kaplan-Meier estimator
method will be employed.

%% Figure1: X-ray light curves for 13 X-ray brightest quasars at $z > 4.5$.

\subsection{Multi-timescale Variability}
\label{sec3.1:multi-timescale variability decomposition via sliding window analysis}

%% Decomposition via Sliding Window Analysis

In our sample, the longest monitoring time on a single source is
about 3.5 years. So, the sample are suitable for studying the
light curve variability across multiple timescales spanning hours
to months. To decouple the short- and intermediate-term
variability components of each quasar, we first compute some
characteristic statistical parameters such as the mean, median and
standard deviation of the flux within an adjacent window of $w$
data points. Specifically, a sliding window width of $w = 10$ is
adopted based on the minimum 14 luminosity measures for a single
source. The variation of the characteristic statistics within
individual windows reveals short-term variability, whereas the
deviation between the mean of each window and the global mean
reflects variability on intermediate timescales. The median
timestamp within each window is adopted as its temporal anchor,
optimizing handling of non-uniformly sampled data.

\begin{figure}[htbp]
\gridline{\fig{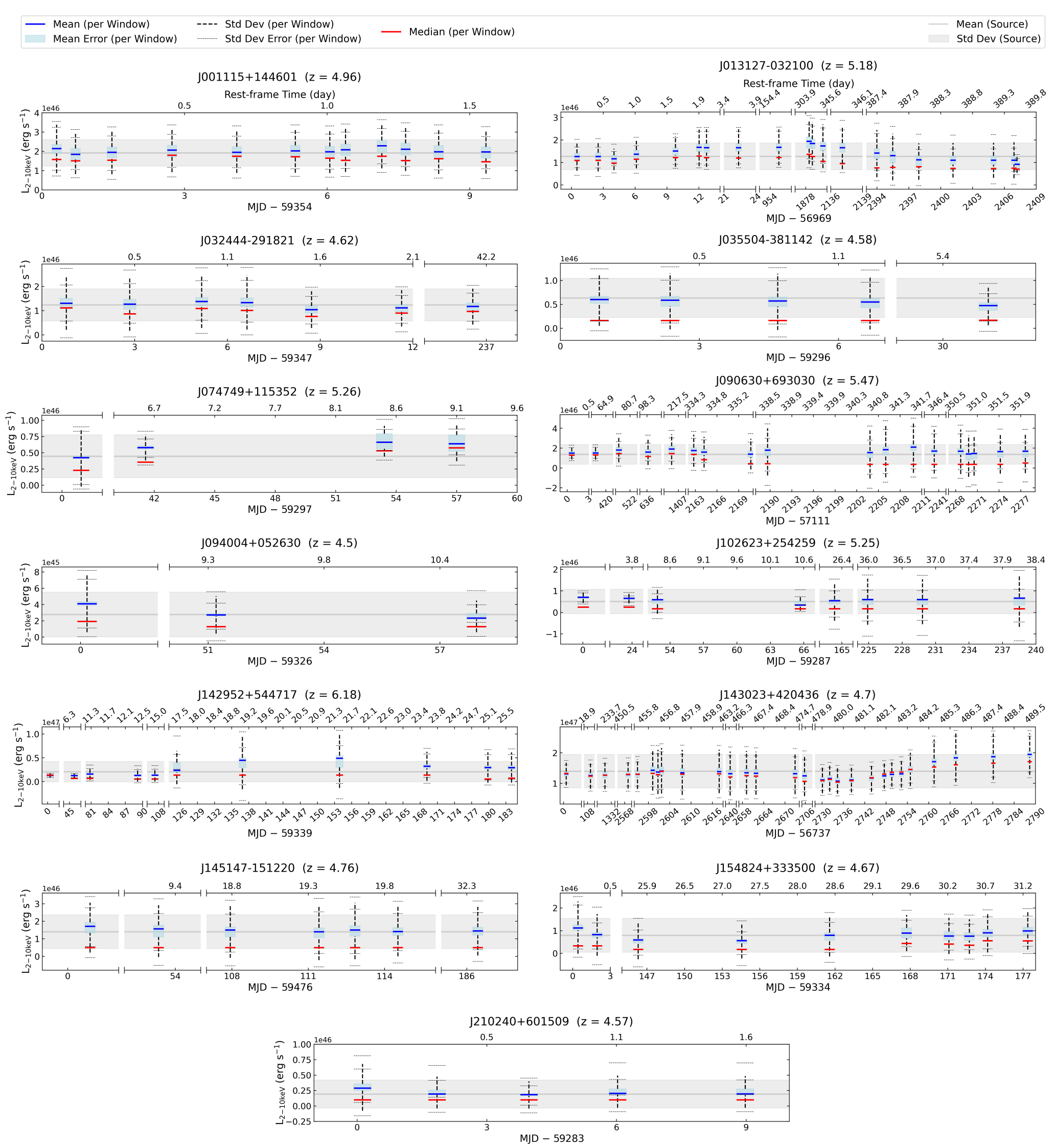}{1\textwidth}{}}
 \caption{ Sliding
window analyses for the 13 quasars. Each dumbbell-shaped marker,
represents a temporal window, including ten temporally adjacent
luminosity measures (including upper limits) in the light curve.
Different colors correspond to different statistics parameters in
the window. The mean luminosity with $\pm\,1\sigma$ uncertainty is
shown in blue. The median luminosity is shown in red. The standard
deviation with $\pm\,1\sigma$ uncertainty is shown by black dashed
line. Gray lines and bands denote the global mean and standard
deviation for the whole light curve. }
 \label{figure2}
\end{figure}

Figure \ref{figure2} illustrates our results of sliding window
analysis for all the 13 sources. Each dumbbell-shaped element
represents a window, with blue lines and shaded regions denoting
window means and uncertainties. Here, the red lines indicate the
medians and the black dashed lines correspond to the standard
deviations. Gray lines and bands display global mean and standard
deviation for the whole light curve. The window statistics
parameters and uncertainties are calculated by using the
Kaplan-Meier estimator method combined with Monte Carlo
simulations (Section \ref{sec2.2:Kaplan-Meier estimator}). Note
that those windows with excessive upper limits are excluded,
resulting in asymmetric distributions of dumbbell elements
relative to global means.

%% Figure2: Sliding window analyses for 13 X-ray brightest quasars at $z > 4.5$.

\subsubsection{Intermediate-term Variability}
\label{sec3.1.1:intermediate-term variability (weeks-to-months)}

Among the 13 quasars, five sources (J032444-291821,
J102623+254259, J142952+544717, J145147-151220, J154824+333500)
were observed for a rest-frame time of $\sim 1$-month. Three
sources (J013127-032100, J090630+693030, J143023+420436) were
monitored for $\sim1$-year (rest-frame). Remaining sources were
only observed for a few days of rest-frame time. In Figure
\ref{figure2}, the comparison between window means (blue lines)
and global means (gray lines) shows the quasars have
intermediate-term variabilities. We see that at intermediate-term
of weeks-to-months, these sources mainly display oscillatory
fluctuations about the global mean, indicating an absence of
secular trends. Such weeks-to-months variabilities could come from
thermal instabilities of accretion disks
\citep{2002MNRAS.332..231U, 2009MNRAS.397.2004A} or coronal
geometry evolution \citep{2016MNRAS.462..511K}. They could also
result from propagating shocks excited by jets
\citep{2016MNRAS.455.3526H}.

\subsubsection{Short-term Variability}
\label{sec3.1.2:short-term variability (hours-to-days)}

All dumbbell elements in Figure \ref{figure2} correspond to
hours-to-days variability. The ratio of mean deviations (window
vs. global) to window standard deviations demonstrates that
short-term variability dominates the X-ray light curve.
Heterogeneity in mean-median-standard deviation relationships
across windows suggests stochastic nature of hours-to-days
variability, particularly evident in the three sources with long
monitoring (J013127-032100, J090630+693030, J143023+420436).
Hours-to-days variability reflects the emission from the regions
very close to the SMBHs \citep{2014A&ARv..22...72U}. It could come
from local magnetic activities or plasma instabilities in the
coronae, or from rapid spectral evolution of electrons in the jet
base.

\subsection{Comparative Study of the Quasars}
\label{sec3.2:comparative analysis of X-ray variability across sample quasars}

\begin{figure}[htbp]
\gridline{\fig{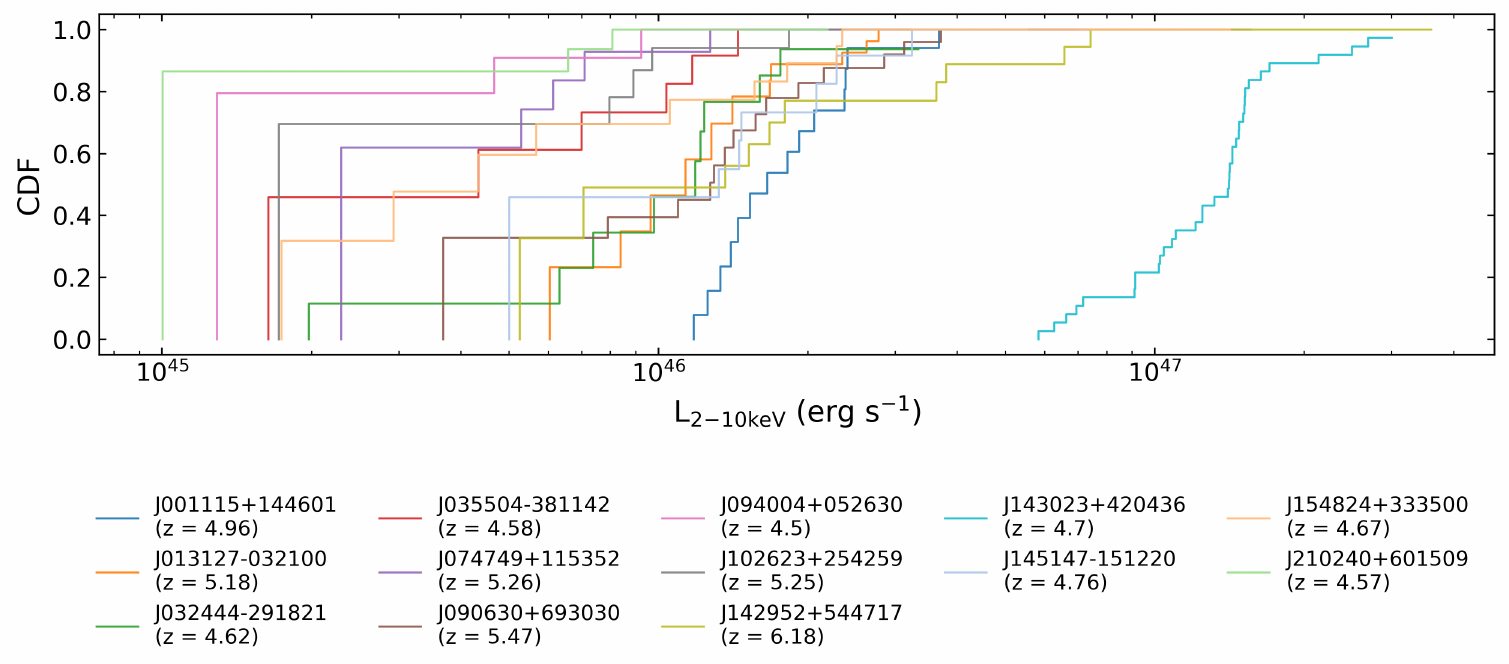}{1\textwidth}{}}
 \caption{ CDF of
rest-frame 2 -- 10 keV luminosity for the 13 quasars in our
sample. The luminosity CDF is calculated by using the Kaplan-Meier
estimator method, which taking into account the upper-limit data
points. }
\label{figure3}
\end{figure}

Using the Kaplan-Meier estimator, we have also calculated the CDF
of the rest-frame 2 -- 10 keV X-ray luminosity for the 13 quasars
in our sample, incorporating the upper limit measurements. The
results are shown in Figure \ref{figure3}. We see that the
luminosity distributions vary across sources, with the ratio
between the maximum and minimum luminosities ranging from $\sim 5$
to $\sim 100$ for most sources. The largest luminosity variation
(a factor of $\sim120$) is observed in J094004+052630. While
J142952+544717, the highest-redshift source in our sample, does
not exhibit the highest median luminosity, its distribution
extends to ${10}^{47}\ \rm{erg\ {s}^{-1}}$. For the source with
the highest median luminosity, J143023+420436, its power is $\sim
{10}^{47}\ \rm{erg\ {s}^{-1}}$ during most of the observations.
The high luminosity is likely due to its exceptionally large
Lorentz factor of the relativistic jet. In fact, this blazar
exhibits the highest Lorentz factor and apparent proper motion
among high-redshift blazars \citep{2020SciBu..65..525Z}, which
also renders it a luminous gamma-ray source
\citep{2018ApJ...865L..17L}. Given its relatively lower redshift
($z = 4.7$) in the high-redshift sample, J143023+420436 has the
highest number of firm X-ray detections (Figure \ref{figure1}),
making it a valuable target for multi-wavelength follow-up.

%% Figure3: CDF of rest-frame 2--10 keV luminosity measures for 13 X-ray brightest quasars at $z > 4.5$.

\begin{figure}[htbp]
\gridline{\fig{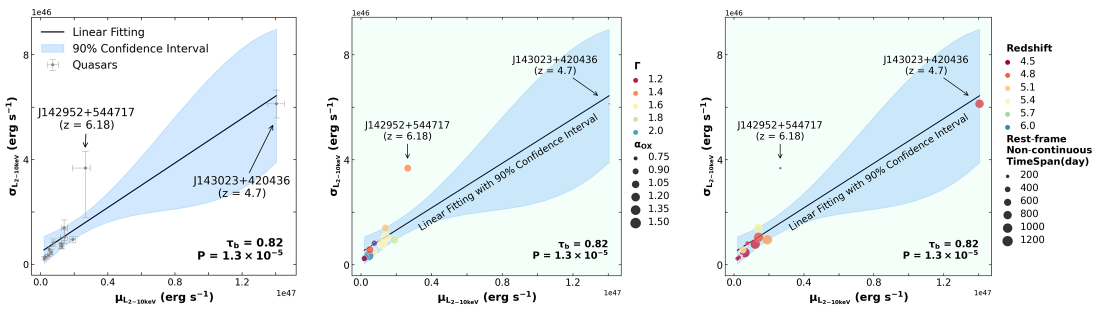}{1\textwidth}{}}
 \caption{$\sigma$-$\mu$ relation of high-redshift AGNs. (a) The global
means ($\mu_{L_{\rm{2\!\text{--}\!10keV}}}$) versus the global
standard deviations ($\sigma_{L_{\rm{2\!\text{--}\!10keV}}}$) of
rest-frame 2 -- 10 keV luminosity for the 13 quasars. The data
points can be well fitted with a linear function (solid line),
with the 90\% confidence interval illustrated as shaded region.
The solid line has a slope of 0.42. The correlation is strong,
which exhibits a Kendall\textquoteright s $\tau_{\rm{b}}$
coefficient of 0.85 and a $P$-value of $1.3\times10^{-5}$ when
tested by using the non-parametric method. (b) The $\sigma$-$\mu$
relation plotted with the X-ray photon index $\Gamma$ color-coded
and the optical-to-X-ray spectral slope size-scaled. (c) The
$\sigma$-$\mu$ relation plotted with the redshift color-coded and
the rest-frame observing span size-scaled. }
 \label{figure4}
\end{figure}

Figure \ref{figure4} plots the global means of luminosity
($\mu_{L_{\rm{2\!\text{--}\!10keV}}}$) versus the global standard
deviations ($\sigma_{L_{\rm{2\!\text{--}\!10keV}}}$) of our
sample. From Figure \ref{figure4}(a), we see that there is a
strong linear correlation between
$\mu_{L_{\rm{2\!\text{--}\!10keV}}}$ and
$\sigma_{L_{\rm{2\!\text{--}\!10keV}}}$.  We have tested the
correlation with the Kendall\textquoteright s $\tau_{\rm{b}}$
method \citep{Kendall1938, Kendall1945, Kendall1970,
2023ApJS..269...17H}. It is found that this correlation is
statistically significant, with a Kendall\textquoteright s
$\tau_{\rm{b}}$ coefficient of 0.85 and a $P$-value of
$1.3\times10^{-5}$.

No correlation is found between
$\mu_{L_{\rm{2\!\text{--}\!10keV}}}$/$\sigma_{L_{\rm{2\!\text{--}\!10keV}}}$
and the X-ray photon index ($\Gamma$; Figure \ref{figure4}b),
implying that the variability is independent of the radiation
mechanism (ICS in coronae or SSC in jets). Similarly, the lack of
correlation with the optical-to-X-ray spectral slope
$\alpha_{\rm{OX}}$  (Figure \ref{figure4}b) suggests that the
variability is governed by localized stochastic processes rather
than large scale accretion processes. Figure \ref{figure4}(c)
further shows that
$\mu_{L_{\rm{2\!\text{--}\!10keV}}}$/$\sigma_{L_{\rm{2\!\text{--}\!10keV}}}$
are independent of redshift and rest-frame monitoring duration,
confirming that the $\sigma$-$\mu$ relation is intrinsic and
unaffected by observational biases. Table \ref{table1} reinforces
this conclusion, as the relation persists across sources with
diverse SMBH masses, pointing to a small-scale physical driver.

This $\sigma$-$\mu$ relation mirrors the well-established
``rms-flux relation'' seen in individual accreting objects (e.g.,
low-redshift AGNs and Galactic black hole X-ray binaries) spanning
over several orders of magnitude in luminosity
\citep{2004MNRAS.348..783M, 2012MNRAS.421.2854S}. However, our
result extends this scaling to an ensemble of distinct
high-redshift quasars. Despite the enhanced accretion rates in the
early universe \citep{2010AJ....140..546W, 2010A&ARv..18..279V,
2015MNRAS.454.3771P, 2015MNRAS.450.1349K, 2020ApJ...897L..14Y},
the similarity of variability statistics across redshifts implies
a possible general mechanism decoupled from large-scale accretion
properties.

As mentioned in Section \ref{sec3.1:multi-timescale variability
decomposition via sliding window analysis}, the X-ray variability
in our sample is dominated by short-term variability originating
near the SMBH. A plausible explanation involves magnetically
driven processes (e.g., magnetic reconnection in the coronae or
jet base), which are natural outcomes of AGN magnetospheric
dynamics \citep{2009ApJ...704...38S, 2011PhPl...18k1207J,
2015MNRAS.450..183S}. If the energy distribution of reconnection
events follows a power law ($\rm{d}N/\rm{d}E \propto E^{-\zeta}$),
the scale-invariance inherent to power-law statistics would
naturally produce the observed $\sigma$-$\mu$ relation. This is
analogous to solar flares, where reconnection energies exhibit
well-documented power-law distributions
\citep{1993SoPh..143..275C, 2000ApJ...535.1047A}, which could be
explained by self-organized criticality
\citep{1991ApJ...380L..89L, 2001SoPh..203..321C,
2022MNRAS.512.1567W} or cascading fragmentation in magnetic
reconnection \citep{2011ApJ...737...24B}.

The prevailing model for the rms-flux relation in X-ray
variability of low redshift AGNs invokes thermal fluctuations in
the accretion disk associated with viscosity
\citep{1997MNRAS.292..679L, 2005MNRAS.359..345U}. However, this
interpretation
%% lacks direct observational validation due to current instrumental resolution
%% limitations in spatially resolving localized thermal disk fluctuations. The
%% model proposed by \citet{1997MNRAS.292..679L}
can mainly account for intermediate-to-long-timescale variability,
as evidenced by the $p(f) \propto f^{-1-4/3b}$ power-law spectral
density (PSD) profile \citep{1997MNRAS.292..679L}.
%% These limitations prevent definitive discrimination between accretion disk
%% fluctuation models and alternative mechanisms.
Instead, short-term variability is more likely due to magnetic
reconnection processes, with the timescale governed by Alfv\'{e}n
waves, i.e. $\tau_{\rm{A}} \sim L_{\rm{A}}/v_{\rm{A}}$, where
$L_{\rm{A}}$ is the current sheet length and $v_{\rm{A}} =
B/\sqrt{\mu_0 \rho}$ is the Alfv\'{e}n velocity. Here $B$ is the
magnetic field strength, $\mu_0$ is the vacuum permeability, and
$\rho$ is the plasma density. For reconnection occurring in
compact regions in the coronae or near the jet base, we have $L
\sim 10 R_{\rm{g}} \approx 10^{13}\,\rm{cm}$, where $R_{\rm{g}}$
is the SMBH gravitational radius. Taking moderate magnetic fields
($B \sim 10^2\,\rm{G}$) and typical electron densities
($n_{\rm{e}} \sim 10^8\,{\rm{cm}}^{-3}$), the derived Alfv\'{e}n
timescale $\tau_{\rm{A}}$ aligns well with the hours-to-days
timescale of short-term X-ray variability. If magnetic
reconnection occurs at a smaller scale, the timescale can be even
shorter.

A universal X-ray variability mechanism across high- and
low-redshift AGNs hinges on whether short-timescale fluctuations
dominate low-$z$ AGN X-ray variabilities. Current analyses reveal
that the PSD of low-$z$ AGNs is not high-frequency-dominated
\citep{2012A&A...544A..80G}. However, stochastic short-timescale
variability, which lacks discrete characteristic frequencies, is
often poorly resolved in PSDs due to non-uniform sampling
artifacts. Conversely, the sliding-window approach can more
effectively decouple these rapid variations. Thus, while observed
intermediate-to-long-timescale X-ray variability in low-$z$ AGN
aligns well with thermal fluctuation models, underlying yet
potentially dominant short-timescale components may remain
obscured. The sliding-window approach method may potentially be
applied to low-$z$ AGNs to reveal their short-timescale
variability.

The slope of the $\sigma$-$\mu$ relation of our high-redshift
quasars is 0.42. It is steeper than that of the low-redshift AGN
rms-flux relation, which is 0.22 \citep{2004MNRAS.348..783M}. If
magnetic reconnection mechanism does dominate the X-ray
variability, such a discrepancy may reflect a stronger magnetic
field or higher reconnection event density in high-redshift AGNs,
although observational selection effects (e.g., luminosity-limited
sampling) may also play a role.

%% Figure4: $\sigma$-$\mu$ relation of high-redshift AGNs.

\subsection{Periodicity Analysis}
\label{sec3.3:periodicity analysis with Lomb-Scargle periodograms}

\begin{figure}[htbp]
\gridline{\fig{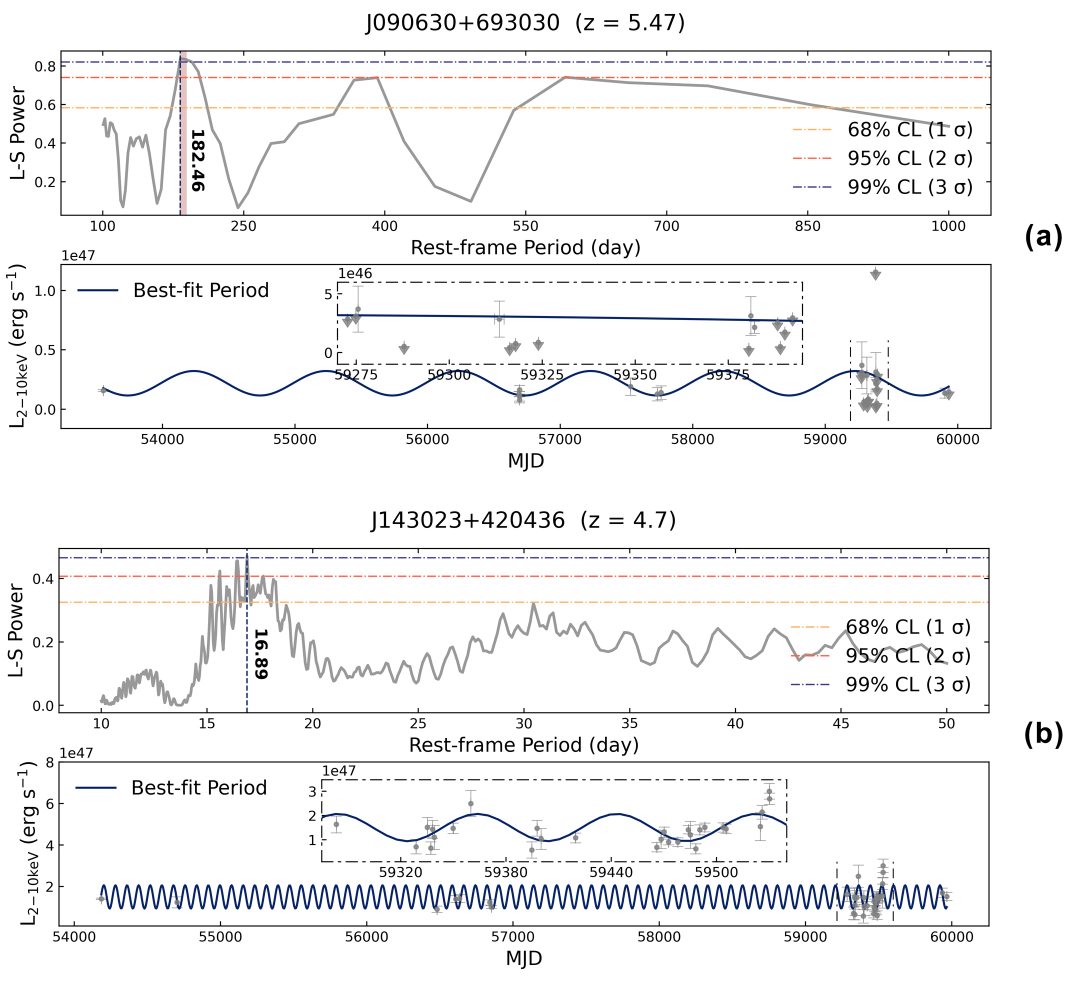}{1\textwidth}{}}
 \caption{Periodicities of J090630+693030 and J143023+420436. (a) The upper panel shows the
Lomb-Scargle periodogram of J090630+693030. The horizontal lines
indicate the confidence levels. The lower panel compares the
best-fit light curve with the observed data points. (b) The upper
panel shows the Lomb-Scargle periodogram of J143023+420436, while
the lower panel compares the best-fit light curve with the
observed data points. }
 \label{figure5}
\end{figure}

We have performed periodicity analysis by using the Lomb-Scargle
periodogram \citep{1976Ap&SS..39..447L, 1982ApJ...263..835S,
2018ApJS..236...16V} for the sources in our sample with sufficient
firm detections. Significant periodic signals exceeding $3\sigma$
confidence level were identified in two sources. J090630+693030
exhibits a rest-frame period of 182.46 days (Figure
\ref{figure5}a, upper panel), though the light curve shows poor
alignment between the best-fit period and the upper-limit
measurements (Figure \ref{figure5}a, lower panel). J143023+420436
displays a rest-frame period of 16.89 days with a better match
between the best-fit model and the luminosity measurements (Figure
\ref{figure5}b, lower panel). Note that J143023+420436 is a radio
and gamma-ray blazar hosting the most relativistic jet among
currently known high-redshift blazars, making it an important
target for multiwavelength follow-up.

Furthermore, discrete magnetic islands formed during the cascading
reconnection in a current sheet may merge during their evolution.
The hierarchical merging of these plasmoids may explain the
observed $\sim 10$-to-100-day periodicity in J090630+693030 and
J143023+420436, since the timescale of global evolution of
plasmoid chains is longer than that of individual reconnection
event \citep{2012PhPl...19d2303L}.

%% Figure5: Periodicities of J090630+693030 and J143023+420436.

\clearpage

%----------------------------------------------------------------------
%                               Conclusions
%----------------------------------------------------------------------

\section{Conclusions}
\label{sec4:conclusions}

Combining the recent Swift observational data of 13 bright
high-redshift X-ray quasars at $z > 4.5$ with archival Swift data
and published history flux data obtained by other satellites, we
systematically investigated the variability and underlying physics
of these sources. The Kaplan-Meier estimator method was employed
to incorporate the observed upper limits of luminosities. It is
found that these high-redshift quasars exhibit X-ray variability
on both short-term (hours-to-days) and intermediate-term
(weeks-to-months) timescales in the rest frame. Generally, the
short-term variability dominates the light curve variation on the
amplitude. A linear correlation exists between the global mean
($\mu_{L_{\rm{2\!\text{--}\!10keV}}}$) and standard deviation
($\sigma_{L_{\rm{2\!\text{--}\!10keV}}}$) of X-ray luminosity,
which is irrespective of the X-ray photon index and
optical-to-X-ray spectral slope, strongly favoring localized
stochastic processes. It is argued that magnetic reconnection in
the coronae or near the jet base could naturally act as the main
radiation mechanism. The timescale of such reconnection process
satisfactorily matches the observed short-term variability, while
the scale-invariant power-law energy distribution inherent to
magnetic reconnection naturally explains the $\sigma$-$\mu$
relation.

The $\sigma$-$\mu$ relation mirrors the well-established "rms-flux
relation" in low-redshift AGNs, indicating that magnetic
reconnection process could potentially act as a universal
mechanism driving short-timescale X-ray variabilities in AGNs,
independent of accretion rate differences between high- and
low-redshift AGNs. The steeper slope of the $\sigma$-$\mu$
relation in high-redshift AGNs ($\sim 0.42$ vs. $\sim 0.22$ for
low-redshift sources) hints a stronger magnetic field or higher
reconnection event density in the early universe, though
observational selection effects cannot be completely excluded yet.

The highest-redshift source in our sample, J142952+544717 ($z =
6.18$), exhibits a not conspicuous median luminosity but displays
a luminosity distribution extending to extreme values (${10}^{47}\
\rm{erg\ {s}^{-1}}$). In contrast, the radio and gamma-ray blazar
J143023+420436 ($z = 4.7$), which hosts the most relativistic jet
among known high-redshift blazars, is dominated in the
high-luminosity regime (${10}^{47}\ \rm{erg\ {s}^{-1}}$ ), making
it an ideal target for multi-wavelength follow-up observations.

Potential periodicity is interestingly found in two sources.
J090630+693030 has a rest-frame period of 182.46 days, and
J143023+420436 has a period of 16.89 days, both at $> 3\sigma$
confidence level. Such a periodicity may arise from the global
evolution of plasmoid chains, in which magnetic islands formed
during reconnection in current sheets may merge successively.
Future multi-wavelength observations, particularly X-ray
polarimetry (e.g., with the Imaging X-ray Polarimetry Explorer
[IXPE]) and time-resolved spectroscopy during flux enhancements,
are essential to probe the role of magnetic reconnection in AGN
variability.

%----------------------------------------------------------------------
%                             Acknowledgements
%----------------------------------------------------------------------

\section*{Acknowledgements}
This study is supported by the National Natural Science Foundation
of China (Grant Nos. 12233002, 12447179),
by National SKA Program of China No. 2020SKA0120300,
by the National Key R\&D Program of China (2021YFA0718500).
YFH also acknowledges the support from the Xinjiang Tianchi Program.

We acknowledge the X-ray data from the Swift public database,
provided at
\href{https://heasarc.gsfc.nasa.gov/cgi-bin/W3Browse/swift.pl}
{https://heasarc.gsfc.nasa.gov/cgi-bin/W3Browse/swift.pl}.

%----------------------------------------------------------------------
%                                References
%----------------------------------------------------------------------

\bibliography{references}{}
\bibliographystyle{aasjournal}

\end{CJK*}
\end{document}